# FERMILAB CRYOMODULE TEST STAND RF INTERLOCK SYSTEM*

T. Petersen†, J. S. Diamond, D. McDowell, D. Nicklaus, P. S. Prieto, A. Semenov,
Fermi National Accelerator Laboratory, Batavia, IL 60510, USA

*Abstract*

An interlock system has been designed for the Fermilab Cryo-module Test Stand (CMTS), a test bed for the cryo-modules to be used in the upcoming Linac Coherent Light Source 2 (LCLS-II) project at SLAC. The interlock system features 8 independent subsystems, one per superconducting RF cavity and solid state amplifier (SSA) pair. Each system monitors several devices to detect fault conditions such as arcing in the waveguides or quenching of the SRF system. Additionally each system can detect fault conditions by monitoring the RF power seen at the cavity coupler through a directional coupler. In the event of a fault condition, each system is capable of removing RF signal to the amplifier (via a fast RF switch) as well as turning off the SSA. Additionally, each input signal is available for remote viewing and recording via a Fermilab designed digitizer board and MVME 5500 processor.

## INTRODUCTION

Fermilab has been designated as one of the laboratories to test a number of cryo-modules for SLAC's upcoming Linac Coherent Light Source 2 [1]. In order to provide facilities for this test, an RF interlocks system has been designed (using previous interlock systems as a template). The interlocks are designed to remove the low level RF (LLRF) within 120 nsec from detection of an RF fault. This is done by monitoring a number of signal sources when an abnormal state is detected, a fast GaAs switch interrupts the LLRF delivered to the SSA. All analog signals are digitized at up to 80 MHz with a 16 channel Fermilab-designed board. The RF interlocks will operate in pulse mode first, then move to continuous wave (CW) mode.

## SYSTEM DESCRIPTION

The full RF interlock system consists of 8 SSA-Coupler pairs controlled by dedicated RF interlocks. Each RF interlock system consists of 5 VME 64X boards which contain analog circuits to process the analog signals and an FPGA to communicate via the VME bus with the crate controller. The interlock boards are organized by function as a System Control board, an 8-channel FWR/REFL PWR board, a 6-channel Field Emission Probe board, a 6-channel Photomultiplier board, and a Digitizer board. These boards can be seen (in the order listed above) in Figure 1. In addition to these boards there is a Relay-Latch board that is used to fan out some auxiliary input signals to the 8 subsystems such as vacuum and temperature trips. A PLC also processes IR and RTD sensors from the couplers and provides an input to the System Control board.

The block diagram given below in Figure 2 illustrates a full system of interlock permits each System Controller monitors. Communication between the RF Interlocks and the Accelerator Controls is done through an MVME 5500 processor, which monitors each board for trips and displays these trips through an EPICS GUI. The controller is also used to remotely sets trip limits (each board has a hard, potentiometer set trip limit as well as a software set, DAC trip limit), and transfer digitizer data.

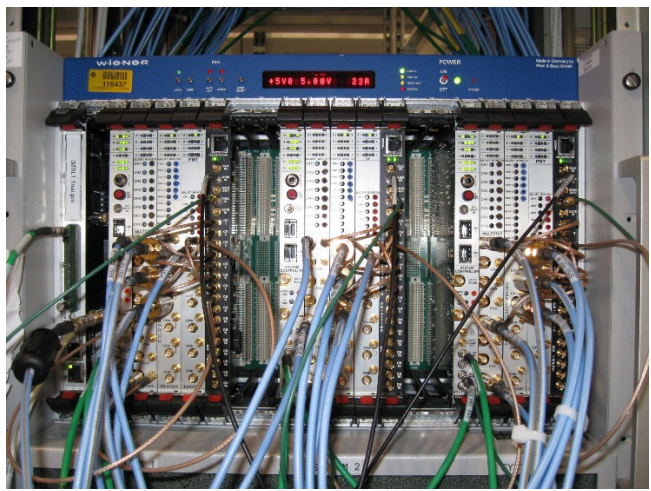

Figure 1: Three full systems in a VME crate with processor and digitizers.

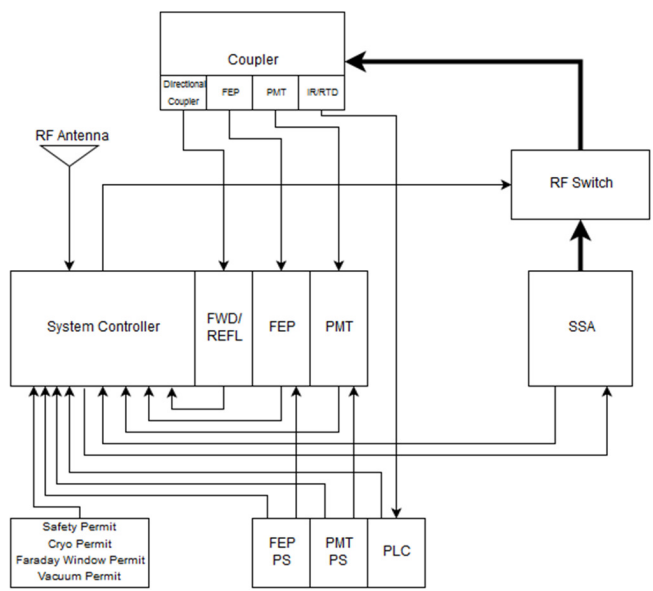

Figure 2: This block diagram details the various signals, permits, and components in each system. Digitizer and MVME 5500 excluded.

___________________________________________
* Work supported by the Fermi National Accelerator laboratory, operated by Fermi Research Alliance LLC, under contract
No. DE-AC02-07CH11359 with the US Department of Energy
† tpeterse@fnal.gov

### System Control Board

The System Control board AND's all digital status signals (active permit high) local interlock boards as well as TTL and contact signals from the PLC, SSA, and Relay-Latch board. The board then generates an enable to the SSA and GaAs fast RF switch allowing LLRF to flow to the SSA. Additionally, one of the 8 System Control cards monitors a 1.3 GHz RF antenna, which monitors for RF leaks from the SSA's. This antenna permit is common to all 8 systems.

### FWR/REFL PWR Board

This board monitors forward RF power, reflected RF power, and transmitted RF power at the cavity coupler via directional couplers. RF is detected using a fast log-amp which allows for monitoring of the input power over a large frequency range. In CW mode, it is proposed that faults from these signals will be determined by computing the sum power of each system, FWR, REFL, TRANS PWR and system losses. In pulse mode operation trips are set by a combination of DAC and trim-pot setting and a signal discriminator. Cavity fill time trips are disabled through a gate permit that controls when the trip is valid based on this fill time.

### Field Emission Probe Board

The Field Emission Probe (FEP) board monitors three FEPs per SSA-Coupler system. The probes monitor three different temperature regions of the coupler (room temp, 80K, and 4K). The probe floats with respect to the plasma it tries to measure. The probe is biased at +20 V, with plans to increase the voltage to approximately +35 V in an effort to increase sensitivity. The FEP is used to detect any ions generated by plasma. A comparator detects the flow of current and inhibits the RF to SSA when limits are exceeded.

### Photomultiplier Tube Board

The Photomultiplier Tube (PMT) board is used to monitor the output signal of a PMT, a Hamamatsu H6780 mounted on the coupler looking through a quartz window. There is also a PMT available looking at the output of the SSA window. The PMT gain is adjustable. The PMT board also utilizes the dark current of the physical device to ensure that cables are connected.

### Auxiliary Inputs

Each system includes an input from a programmable logic controller (PLC) that monitors two temperature sensors: an infrared (IR) detector, and a resistive temperature detector (RTD).

In addition to monitoring these and previously mentioned devices, each system monitors auxiliary inputs as well. These include a Cryo permit, a Vacuum system permit, temperature sensors on a Faraday window of the Cryomodule, and permits from both FEP and PMT external power supplies. These five signals are fanned out (along with the antenna trip signal previously mentioned) so that they are common to all 8 systems.

### Digitizers

All of the 8 systems include a Fermilab designed 16 channel, 80 MHz digitizer board for monitoring each input signal. Each interlock board has a monitor output for each channel that enables the user to see the response while running. The system is designed with two types of monitoring capabilities.

The first utilizes a circular buffer that keeps a smaller length of data (in time) but with much higher sampling rate. The sampling rate and number of samples stored are variable, allowing the user to adjust the window around the trip condition. This buffer will be read by the processor when a trip occurs, allowing the user to record and view the events that caused the trip in high detail. An example of this can be seen in Figure 3, which shows FEP activity that caused a trip. During this mode the data for each channel can be plotted in real time at 720 Hz.

The second mode decimates the high sampling rate data in order to look at the entirety of the pulse in pulsed mode, with an adjustable sampling rate. This variable sampling rate allows the user to look at the entire pulse even as the pulse width grows during its approach to CW mode.

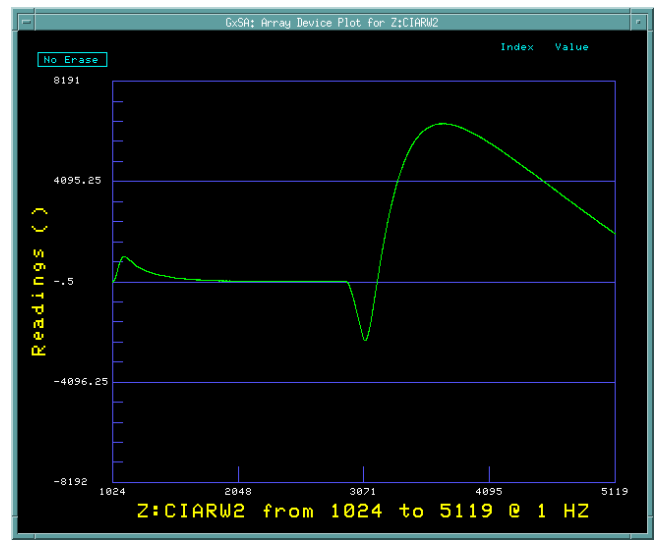

Figure 3: 80 MHz sampling rate circular buffer data showing FEP activity causing a trip.

## USAGE

The interlock systems are remotely monitored, and outfitted so that operation of the interlock system can be done remotely for most situations. This is done via a Synoptic (an EPICS GUI) display window, shown in Figure 4, as well as various ACNET (Fermilab's Accelerator Control system) pages. All of the interlock boards have both analog trip limit settings as well as a digitally set trip limit, which can be set via the Synoptic display.

Additionally each board displays their status, allowing a user to determine which board (and which specific device) caused a trip. This allows the user to determine if the trip requires attention, as well as aiding in the conditioning of the couplers. A trip condition can be reset remotely for all

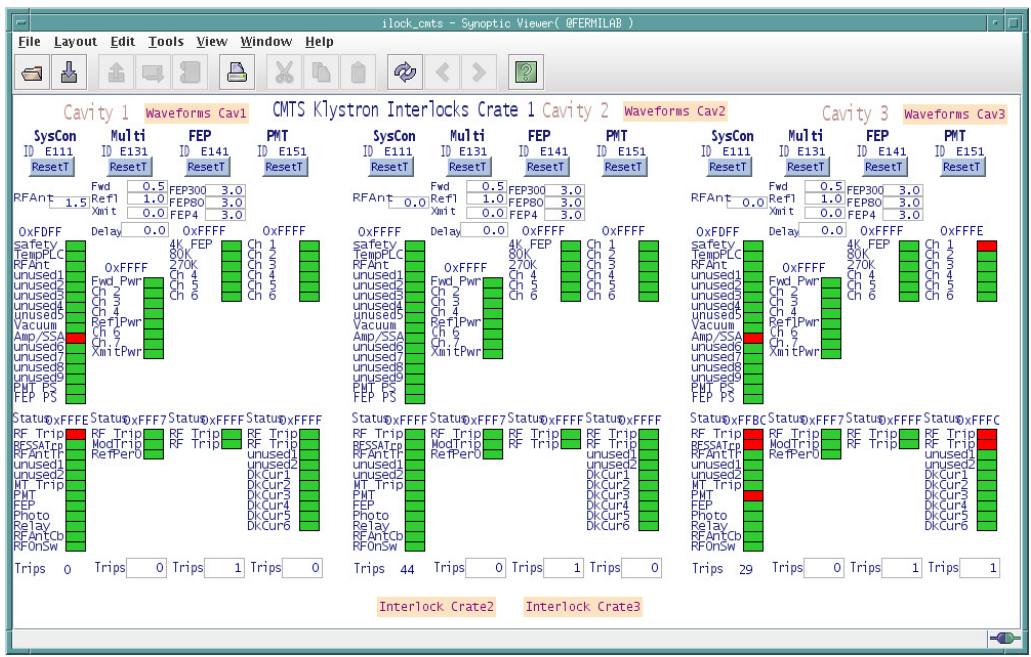

Figure 4: EPICS GUI display for one RF Interlock crate. The three systems are shown in various states, with System 2 permitted to run RF power into the cavity

boards, allowing the user to remotely clear the interlocks and continue running.

The digitizer mode, as well as sampling rate and number of samples taken, is also set remotely via ACNET. For the real time 720 Hz plots the data is plotted in ACNET, while for the pulsed mode and circular buffer readout (following a trip condition) can be accessed via the Synoptic display.

## CONCLUSION

The prototype Cryo-module has been installed and testing has begun. The interlock system has been functioning as expected, allowing the eight systems to be run while monitoring all systems for a trip condition.

## ACKNOWLEDGEMENTS

The authors gratefully acknowledge the work done by the technicians, engineers, and physicists in the Instrumentation department and Controls department.